\documentclass[pdftex,twocolumn,epjc3]{svjour3}

\usepackage{amsfonts}
\usepackage{amsmath}
\usepackage{amssymb,bbold}
\usepackage[normalem]{ulem}

\RequirePackage[T1]{fontenc}
\smartqed
\RequirePackage{graphicx}
\RequirePackage{mathptmx}
\RequirePackage{flushend}
\RequirePackage[numbers,sort&compress]{natbib}
\RequirePackage[colorlinks,citecolor=blue,urlcolor=blue,linkcolor=blue]{hyperref}

\journalname{Eur. Phys. J. C}

\RequirePackage{color}

\begin{document}

\title{Effects of spin on the dynamics of the 2D Dirac oscillator in the
  magnetic cosmic string background}

\author{Fabiano M. Andrade\thanksref{e1,addr1}
        \and
        Edilberto O. Silva\thanksref{e2,addr2}
}

\thankstext{e1}{e-mail: fmandrade@uepg.br}
\thankstext{e2}{e-mail: edilbertoo@gmail.com}

\institute{Departamento de Matem\'{a}tica e Estat\'{i}stica,
           Universidade Estadual de Ponta Grossa,
           84030-900 Ponta Grossa-PR, Brazil\label{addr1}
           \and
           Departamento de F\'{i}sica,
           Universidade Federal do Maranh\~{a}o,
           Campus Universit\'{a}rio do Bacanga,
           65085-580 S\~{a}o Lu\'{i}s-MA, Brazil\label{addr2}
         }

\date{Received: date / Accepted: date}

\maketitle

\begin{abstract}
In this work the dynamics of a 2D Dirac oscillator in the
spacetime of a magnetic cosmic string is considered.
It is shown that earlier approaches to this problem have neglected a
$\delta$ function contribution to the full Hamiltonian, which comes from
the Zeeman interaction.
The inclusion of spin effects leads to results which confirm a modified
dynamics.
Based on the self-adjoint extension method, we determined the most
relevant physical quantities, such as energy spectrum, wave functions
and the self-adjoint extension parameter by applying boundary conditions
allowed by the system.
\end{abstract}

\section{Introduction}
\label{sec:introduction}

The Dirac oscillator is a natural model for studying properties of
physical systems.
This model is based on the dynamics of a harmonic oscillator for
spin-1/2 particles by introducing a nonminimal prescription into free
Dirac equation \cite{JPA.1989.22.817}.
Because it is a exactly solvable model, several investigations have been
developed in the context of this theoretical framework in the last
years.
The interest in this issue appears in different contexts, such as
quantum optics
\cite{AP.2013.331.120,PRA.2008.77.063815,PRA.2007.76.041801},
supersymmetry \cite{JMP.1992.33.1831,PRL.1990.64.1643,PRD.1991.43.544},
nuclear reactions \cite{PRC.2009.80.044607}, Clifford algebra
\cite{PLA.2008.372.2587,JMP.1993.34.4428} and noncommutative space
\cite{JMP.2014.55.032105,IJTP.2010.49.1699}.
Recently, the one-dimensional Dirac oscillator has been verified
experimentally by J. A. Franco-Villafa\~{n}e et al., based on a
tight-binding system \cite{PRL.2013.111.170405}.
A detailed description for the Dirac oscillator is given in
Ref. \cite{Book.1998.Strange} and for other contributions see Refs.
\cite{PLB.2014.731.327,AoP.2013.336.489,PRA.2011.84.32109,
PLA.2005.346.261,AHEP.2013.2013.383957,JPA.2005.38.1747,
JPA.2006.39.5125,PLA.2004.325.21}.

Among the various contexts in which the Dirac oscillator can be
addressed, we refer to the cosmic string, a linear defect that change
the topology of the the medium when viewed globally.
This framework has inspired a great deal of investigation in recent
years. Such works encompass several distinct aspects to investigate the
effects produced by topological defects of this nature
\cite{PRD.2011.83.125025,PRD.2014.89.027702,PLA.2007.361.13,
NPB.1989.328.140,PRL.1989.62.1071,PRD.2012.85.041701,AoP.2013.339.510}.

In this work, we generalize the results in  \cite{PRA.2011.84.32109}
for a 2D Dirac oscillator in the magnetic cosmic string background
showing rigorously how the dynamics of this system is affected when the
effects of spin are taken into account.
Our approach is based on the self-adjoint extension method which is
appropriate to address any system endowed with a singular Hamiltonian
(due to localized field sources or quantum confinement).
We determine the most relevant physical quantities from the present
model, such as energy spectrum, wave functions, and self-adjoint
extension parameter by applying boundary conditions allowed by the
system.

\section{The 2D Dirac oscillator in the magnetic cosmic string
  background}
\label{sec:sec2}

In this section, we study the motion of the particle in the magnetic cosmic
string background.
The cosmic string spacetime with an internal magnetic field is an object
described by the following line element in cylindrical coordinates
($t,r,\varphi,z$):
\begin{equation}
ds^{2}=-dt^{2}+dr^{2}+\alpha ^{2}r^{2}d\varphi^{2}+dz^{2},
\label{eq:metric}
\end{equation}
with $-\infty <(t,z)<\infty $, $r\geq 0$ and $0\leq \varphi \leq 2\pi$.
The parameter $\alpha $ is related to the linear mass density
$\tilde{m}$ of the string by $\alpha =1-4\tilde{m}$ runs in the interval
$(0,1]$ and corresponds to a deficit angle $\gamma =2\pi (1-\alpha )$.
Geometrically, the metric \eqref{eq:metric} corresponds to a Minkowiski
spacetime with a conical singularity \cite{JHEP.2004.2004.16}.

We begin with the Dirac equation in the curved spacetime
(with $\hbar=c=1$):
\begin{equation}\label{eq:dirac}
\left[
  i\gamma^{\mu}(\partial_{\mu}+\Gamma_{\mu})
  -e\gamma^{\mu}A_{\mu}-M
  \right]\Psi=0,
\end{equation}
where $e$ is the charge, $M$ is mass of the particle, $\Psi$ is a
four-component spinorial wave function, and $\Gamma_{\mu}$ is the spinor
affine connection, which is given by
\cite{Book.2003.Carroll,APPB.2010.41.1827}
\begin{equation}\label{eq:conn}
  \Gamma_{\mu}=\frac{1}{4}i\omega_{\mu \bar{a}\bar{b}}
  \sigma_{\bar{a}\bar{b}}=
  -\frac{1}{8}\omega_{\mu\bar{a}\bar{b}}
  \left[ \gamma^{\bar{a}},\gamma^{\bar{b}}\right],
\end{equation}
where $\gamma^{\bar{a}}$ are the standard Dirac matrices in Minkowski
spacetime, and $\omega_{\mu \bar{a}\bar{b}}$ is the spin connection,
given by
\begin{equation}
  \omega_{\mu \bar{a}\bar{b}}=
  \eta_{\bar{a}c}e_{\nu}^{\bar{c}}e_{\bar{b}}^{\sigma}
  \Gamma_{\sigma \mu}^{\nu}-  \eta_{\bar{a}\bar{c}}e_{\nu}^{\bar{c}}
  \partial_{\mu}e_{\bar{b}}^{\nu},  \label{spincn}
\end{equation}
with $(\mu ,\nu )=(0,1,2,3)$ and $(\bar{a},\bar{b})=(0,1,2,3)$;
$\Gamma_{\sigma \mu}^{\nu}$ is the Christoffel symbol,
$\eta^{\bar{a}\bar{b}}$ is the metric tensor, $e_{\bar{a}}^{\nu}$ is the
basis tetrad which will be defined below.
The spin connection \eqref{eq:conn} allows us to construct a local frame
through the basis tetrad which gives the spinors in the curved
spacetime.
Also, the $\gamma^{\mu}$ matrices are the generalized Dirac matrices
defining the covariant Clifford algebra,
\begin{equation}
  \left\{ \gamma^{\mu},\gamma^{\nu}\right\} =2g^{\mu \nu},
  \label{eq:comut}
\end{equation}
and are written in terms of the standard Dirac matrices
$\gamma^{\bar{a}}$ in Minkowski spacetime as
\begin{equation}
  \gamma^{\mu}=e_{\bar{a}}^{\mu}\gamma^{\bar{a}},  \label{eq:diracmathe}
\end{equation}
with
$\gamma^{\bar{a}}=\left( \gamma^{\bar{0}},\gamma^{\bar{\imath}}\right)$,
and
\begin{equation}
  \gamma^{\bar{0}}=
  \left(
    \begin{array}{rr}
      \mathbb{1} & 0 \\
      0 & \mathbb{-1}
    \end{array}
  \right) ,~~~
  \gamma^{\bar{\imath}}=
  \left(
    \begin{array}{cc}
      0 & \sigma^{i} \\
      -\sigma^{i} & 0
    \end{array}
  \right) ,~~~(i=1,2,3)
\end{equation}
where $\sigma^{i}$ are the standard Pauli matrices and $\mathbb{1}$ is
the $2\times 2$ identity matrix.
The basis tetrad $e_{\bar{a}}^{\mu}$ in Eq. \eqref{eq:diracmathe} is
chosen to be \cite{PRA.2011.84.32109}
\begin{equation}
  e_{\bar{a}}^{\mu}=
  \left(
    \begin{array}{cccc}
      1 & 0 & 0 & 0 \\
      0 & \cos \varphi & \sin \varphi & 0 \\
      0 & -\sin \varphi /\alpha r & \cos \varphi /\alpha r & 0 \\
      0 & 0 & 0 & 1
    \end{array}
  \right) ,  \label{eq:tetrad}
\end{equation}
satisfying the condition
\begin{equation}
  e_{\bar{a}}^{\mu}e_{\bar{b}}^{\nu}\eta^{\bar{a}\bar{b}}=g^{\mu \nu}.
\end{equation}
The matrices $\gamma^{\mu}$ in Eq. \eqref{eq:diracmathe} are given more
explicitly as
\begin{subequations}
\begin{align}
  \gamma^{0} = {} &
  e_{0}^{t}\gamma^{\bar{0}}\equiv \gamma^{t},  \label{eq:ga0} \\
  \gamma^{3} = {} &
  e_{\bar{0}}^{z}\gamma^{\bar{0}}\equiv\gamma^{z}, \label{eq:gaz} \\
  \gamma^{1} = {} &
  e_{\bar{a}}^{1}\gamma^{\bar{a}}\equiv \gamma^{r},  \label{eq:ga1} \\
  \gamma^{r} = {}  &
  e_{\bar{0}}^{r}\gamma^{\bar{0}}+e_{\bar{1}}^{r}\gamma^{\bar{1}}+
  e_{\bar{2}}^{r}\gamma^{\bar{2}}=
  \gamma^{\bar{1}}\cos \varphi +\gamma^{\bar{2}}\sin \varphi,   \\
  \gamma^{2} = {} &
  e_{\bar{a}}^{2}\gamma^{\bar{a}}\equiv
  \frac{\gamma^{\varphi}}{\alpha r}, \label{eq:ga2} \\
  \gamma^{\varphi} = {} &
  e_{\bar{0}}^{\varphi}\gamma^{\bar{0}}+
  e_{\bar{1}}^{\varphi}\gamma^{\bar{1}}+e_{\bar{2}}^{\varphi}\gamma^{\bar{2}}=
  -\gamma^{\bar{1}}\sin \varphi +\gamma^{\bar{2}}\cos \varphi .
\end{align}
\end{subequations}
The starting point for the derivation of Eq. \eqref{eq:conn} is that the
curved-space gamma matrices are covariantly constant
\cite{PR.1981.68.189,Book.2003.Carroll,APPB.2010.41.1827}, i.e.,
$\nabla_{\mu}\gamma^{\lambda}=0$ (see \ref{sec:appendixA}).
For the specific basis tetrad \eqref{eq:tetrad}, the connection is found
to be
\begin{equation}
\boldsymbol{\Gamma}=\left( 0,0,\Gamma_{\varphi},0\right),
\label{eq:connec}
\end{equation}
with the non-vanishing element given as
\begin{equation}
  \Gamma_{\varphi}=-\frac{1}{2}\left(1-\alpha \right)
  \gamma_{\bar{1}}\gamma_{\bar{2}}.  \label{eq:gammaphi}
\end{equation}
Details for the calculation of the connection are given in
\ref{sec:appendixB}.

According to the tetrad postulated
\cite{PR.1981.68.189,Book.2012.Lawrie,Book.2003.Carroll,
APPB.2010.41.1827},
the matrices $\gamma^{\bar{a}}$ could be any set of constant Dirac
matrices.
Thus, we are free to choose a representation for the matrices
$\gamma^{\bar{a}}$.
Making use of the symmetry under z translations of the
system, we can reduce the four-component Dirac equation
\eqref{eq:dirac} to a two two-component spinor equations.
To do this, we consider the vector potential $\mathbf{A}$ as being
intrinsically two-dimensional, i.e., it has only two components and
depends on only two spatial coordinates and we take $p_{z}=z=0$.
In this manner, the relevant equation is
\begin{equation}
  \left[
    \tilde{\beta} \boldsymbol{\tilde{\gamma}}\cdot \boldsymbol{\pi}
    +\tilde{\beta} M
  \right] \psi =E\psi ,
  \label{eq:dirac2}
\end{equation}
where $\psi$ is a two-component spinor, and
\begin{equation} \label{eq:meff}
 \boldsymbol{\pi}=
  -i(\boldsymbol{\nabla}_{\alpha}+\boldsymbol{\Gamma})
  -e\mathbf{A},
\end{equation}
is the generalized momentum,
\begin{equation}
  \boldsymbol{\nabla}_{\alpha}=
  \frac{\partial}{\partial r}\mathbf{\hat{r}}+
  \frac{1}{\alpha r}\frac{\partial}{\partial\varphi}
  \hat{\boldsymbol{\varphi}},
\end{equation}
is the gradient operator in polar coordinates, and the
$\tilde{\gamma}^{\bar{a}}$ matrices are given in terms of the Pauli
matrices as
\begin{equation}\label{eq:newmatrices}
  \tilde{\beta} =\tilde{\gamma}^{\bar{0}}=\sigma^{z},\qquad
  \tilde{\beta} \tilde{\gamma}^{\bar{1}}=\sigma^{1},\qquad
  \tilde{\beta} \tilde{\gamma}^{\bar{2}}=s\sigma ^{2},
\end{equation}
where the parameter $s$, which has a value of twice the spin value, can
be introduced to characterizing the two spin states
\cite{PRL.1990.64.503,IJMPA.1991.6.3119}, with $s=+1$ for
spin ``up'' and $s=-1$ for spin ``down'' \cite{EPJC.2014.74.2708}.

In the representation \eqref{eq:newmatrices}, Eq. \eqref{eq:gammaphi}
yields
\begin{equation}\label{eq:newconx}
  \Gamma_{\varphi}=-\frac{1}{2}
  \left( 1-\alpha \right) i\sigma _{2}\left(-is\sigma _{1}\right)=
  i\frac{\left( 1-\alpha \right) }{2}s\sigma^{z}.
\end{equation}

The magnetic vector potential in polar coordinates in the Coulomb
gauge is chosen to be
\begin{equation}
  e\mathbf{A}=-\frac{\phi}{\alpha r}\hat{\boldsymbol{\varphi}},
\label{eq:vectora}
\end{equation}
where $\phi =\Phi /\Phi_{0}$ is the flux parameter with
$\Phi_{0}=2\pi/e$
($(\hat{\mathbf{r}},\hat{\boldsymbol{\varphi}})$
denote the unit vectors in polar coordinates.)
This choice for the vector potential gives a magnetic flux tube, in the
background space described by the metric \eqref{eq:metric}, coinciding
with the cosmic string and with the magnetic field strength given by
\begin{equation}\label{eq:vectorb}
  e B=-\frac{\phi}{\alpha}\frac{\delta(r)}{r}.
\end{equation}
Note that, in the limit as $\alpha \rightarrow 1$, we obtain the magnetic
field in Euclidean space.

The 2D Dirac oscillator is introduced by the non-minimal substitution
\cite{JPA.1989.22.817},
\begin{equation}
  \frac{1}{i}\boldsymbol{\nabla}_{\alpha}\rightarrow
  \frac{1}{i}\boldsymbol{\nabla}_{\alpha}
  -iM\omega \beta \mathbf{r},
\end{equation}
where $\mathbf{r}$ is the position vector and $\omega$ the frequency of the
oscillator (for a comprehensive discussion of the Dirac oscillator see Ref.
\cite{Book.1998.Strange}).
In this case, Eq. \eqref{eq:dirac2} reads
\begin{equation}
  \left[
    \boldsymbol{\alpha}\cdot
    (\boldsymbol{\pi}-iM\omega \beta \mathbf{r})
    +\beta M
  \right] \psi =E\psi .  \label{eq:dirac3}
\end{equation}

The second order equation implied by Eq. \eqref{eq:dirac3} is obtained
by applying the matrix operator
\begin{equation}
\left[\beta M+E+\boldsymbol{\alpha}\cdot
(\boldsymbol{\pi}-iM\omega \beta \mathbf{r})\right].
\end{equation}
So, one finds
\begin{equation}  \label{eq:eqsb1}
  (E^{2}-M^{2}) \psi =
  [\boldsymbol{\alpha}\cdot(\boldsymbol{\pi}-iM\omega\beta\mathbf{r})]
  [\boldsymbol{\alpha}\cdot(\boldsymbol{\pi}-iM\omega\beta\mathbf{r})]
  \psi.
\end{equation}
Inserting  Eqs. \eqref{eq:vectorb}, \eqref{eq:vectora} and the
expression for  $\Gamma_{\varphi}$ in \eqref{eq:newconx} into
Eq. \eqref{eq:eqsb1}, one obtains
\begin{equation}\label{eq:edod}
  (E^{2}-M^{2})\psi=H\psi,
\end{equation}
where
\begin{align}\label{eq:heff}
  H = {} &
  \left[
    -i\boldsymbol{\nabla}_{\alpha}
    +\left(
      \frac{\phi}{\alpha}+\frac{1-\alpha}{2\alpha}s\sigma^{z}
    \right)\frac{1}{r}
    \hat{\boldsymbol{\varphi}}
  \right]^{2}\nonumber \\
  &-2M\omega
  \left[\sigma^{z}+
    s\left(\frac{1}{i\alpha}\frac{\partial}{\partial\varphi}
    +\frac{\phi}{\alpha}+\frac{1-\alpha}{2\alpha}s\sigma^{z}
    \right)
  \right]
  \nonumber \\
  & +M^{2}\omega^{2}r^{2}
  +\frac{\phi s}{\alpha}\frac{\delta(r)}{r}\sigma^{z}.
\end{align}
In Eq. \eqref{eq:heff}, the quantity
\begin{equation}
\frac{\phi}{\alpha}+\frac{1-\alpha}{2\alpha}s\sigma^{z},
\end{equation}
contributes to the term which depends explicitly on the spin of the
particle.
The first term is the contribution due to the magnetic flux while the
second is due to the spin connection.
Note that, by making $\alpha=1$ (flat spacetime) and $\phi=0$
(absence of a magnetic field) in Eq. \eqref{eq:edod}, we obtain, for the
planar case, the 2D Dirac oscillator as proposed by Moshinsky and
Szczepaniak \cite{JPA.1989.22.817} and discussed in
\ref{sec:appendixC}.

Making use of the underlying rotational symmetry we can express the
two-component spinor as
\begin{equation}\label{eq:soloc}
  \psi(r,\varphi)=
   \left(
    \begin{array}{c}
      \psi_{1}\\
      \psi_{2}
    \end{array}
  \right)=
  \left(
    \begin{array}{c}
      f_{m}(r)\;e^{i m \varphi} \\
      g_{m}(r)\;e^{i(m +s)\varphi}
    \end{array}
  \right),
\end{equation}
with $m\in \mathbb{Z}$.
By replacing Eq. \eqref{eq:soloc} into Eq. \eqref{eq:edod}, we obtain
the radial equation for $f_{m}(r)$
\begin{equation}\label{eq:eigen}
  H f_{m}(r) = k^{2} f_{m}(r),
\end{equation}
where
\begin{equation}
k^{2}=E^{2}-M^{2}+2 M\omega(s j + 1),
\end{equation}
\begin{equation}
  j=\frac{1}{\alpha}
  \left(
    m+\phi+\frac{1-\alpha}{2}s
  \right),
\end{equation}
\begin{equation}\label{eq:hfull}
  H=H_{0} +\frac{\phi s}{\alpha}\frac{\delta(r)}{r},
\end{equation}
and
\begin{equation}\label{eq:hzero}
  H_{0}=
    -\frac{d^{2}}{dr^{2}}-\frac{1}{r}\frac{d}{dr}
    +\frac{j^2}{r^{2}}
    +M^2\omega^{2}r^{2}.
\end{equation}
The Hamiltonian in Eq. \eqref{eq:hfull} governs the dynamics of a Dirac
oscillator in a magnetic cosmic string background, i.e., a Dirac
oscillator problem in the presence of the Aharonov-Bohm effect in a
conical spacetime.
The presence of a two-dimension $\delta$ interaction in the radial
Hamiltonian $H$, which is singular at the origin, makes the problem
more complicated to solve.
The most adequate manner to address this kind of point interaction
potential is by making use of the self-adjoint extension approach
\cite{Book.2004.Albeverio,JMP.1985.26.2520}.
This is the method adopted in this work and discussed in the next
section.

\section{Self-adjoint extension analysis}
\label{sec:selfae}

In this section, we review some concepts on the self-adjoint extension
approach.
An operator $\mathcal{O}$, with domain $\mathcal{D}({\mathcal{O}})$, is
said to be self-adjoint if and only if
$\mathcal{O}=\mathcal{O}^{\dagger}$ and
$\mathcal{D}(\mathcal{O})=\mathcal{D}(\mathcal{O}^{\dagger})$,
$\mathcal{O}^{\dagger}$ being the adjoint of operator $\mathcal{O}$.
For smooth functions, $\xi \in C_{0}^{\infty}(\mathbb{R}^2)$ with
$\xi(0)=0$, we should have $H \xi =H_{0} \xi$, and it is
possible to interpret the Hamiltonian \eqref{eq:hfull} as a
self-adjoint extension of $H_{0}|_{C_{0}^{\infty}(\mathbb{R}^{2}/\{0\})}$
\cite{crll.1987.380.87,JMP.1998.39.47,LMP.1998.43.43}.
The self-adojint extension approach consists, essentially, in extending
the domain of $\mathcal{D}(\mathcal{O})$ in order to match
$\mathcal{D}(\mathcal{O}^{\dagger})$.
From the theory of symmetric operators, it is a well-known fact that the
symmetric radial operator $H_{0}$ is essentially self-adjoint for
$|j|\geq 1$, while for $|j|< 1$ it admits an
one-parameter family of self-adjoint extensions
\cite{Book.1975.Reed.II}, $H_{0,\lambda_m}$, where $\lambda_{m}$ is the
self-adjoint extension parameter.
To characterize this family, we will use the approach in
\cite{JMP.1985.26.2520,Book.2004.Albeverio}, which is based on the
boundary conditions at the origin. 
All the self-adjoint extensions $H_{0,\lambda_{m}}$  of $H_{0}$ are
parametrized by the boundary condition at the origin
\begin{equation}
\label{eq:bc}
  \Omega_{0}=\lambda_{m} \Omega_{1},
\end{equation}
with
\begin{align}
  \Omega_{0} = {} & \lim_{r\rightarrow 0^{+}}r^{|j|} f_{m}(r),
  \\
  \Omega_{1} = {} & \lim_{r\rightarrow 0^{+}}\frac{1}{r^{|j|}}
  \left[
    f_{m}(r)-\Omega_{0}\frac{1}{r^{|j|}}
  \right],
\end{align}
where $\lambda_{m}\in \mathbb{R}$ is the self-adjoint extension
parameter.
For $\lambda_{m}=0$, we have the free Hamiltonian (without the
$\delta$ function)  with regular wave functions at the origin,
and for $\lambda_{m}\neq 0$  the boundary condition in
Eq. \eqref{eq:bc}
permit an $r^{-|j|}$ singularity in the wave functions at
the origin.

\section{The bound state energy and wave function}
\label{sec:tbse}

In this section, we determine the energy spectrum for the Dirac
oscillator in the cosmic string background by solving
Eq. \eqref{eq:eigen}.
For $r\neq 0$, the equation for the component $f_{m}(r)$ can be transformed
by the variable change $\rho=M\omega r^2$ resulting in
\begin{equation}\label{eq:edofrho}
  \rho f_{m}''(\rho)+f_{m}'(\rho)
  -\left(
    \frac{{j}^2}{4\rho}+\frac{\rho}{4} -\frac{k^2}{4\gamma}
  \right)f_{m}(\rho)=0,
\end{equation}
with $\gamma=M\omega$.
Due to the boundary condition in Eq. \eqref{eq:bc}, we seek for regular
and irregular solutions for Eq. \eqref{eq:edofrho}.
Studying the asymptotic limits of Eq. \eqref{eq:edofrho} leads us to the
following regular ($+$) (irregular ($-$)) solution:
\begin{equation}\label{eq:frho}
  f_{m}(\rho)=\rho^{\pm|j|/2}e^{-\rho/2} F(\rho).
\end{equation}
With this, Eq. \eqref{eq:edofrho} is rewritten as
\begin{multline}\label{eq:edoMrho}
  \rho F''(\rho)
  +\left(1\pm |j|-\rho\right) F'(\rho) \\
  -\left(
    \frac{1\pm |j|}{2}-\frac{k^2}{4\gamma}
  \right)F(\rho)=0.
\end{multline}
Equation \eqref{eq:edofrho} is of the confluent hypergeometric equation
type
\begin{equation}
  z F''(z)+(b-z) F'(z)-a F(z)=0.
\end{equation}
In this manner, the general solution for Eq. \eqref{eq:edofrho} is
\begin{align}\label{eq:general_sol_2_HO}
  f_{m}(r)
  = {} &
  a_{m} \rho^{|j|/2} e^{-\rho/2}\;
  F\left(d_{+},1+|j|,\rho\right)\nonumber \\
  & + b_{m}\rho^{-|j|/2}e^{-\rho/2}\;
  F\left(d_{-},1-|j|,\rho\right),
\end{align}
with
\begin{equation}
  d_{\pm}=\frac{1 \pm |j|}{2}-\frac{k^2}{4 \gamma}.
\end{equation}
In Eq. \eqref{eq:general_sol_2_HO}, $F(a,b,z)$ is the
confluent hypergeometric function of the first kind
\cite{Book.1972.Abramowitz} and $a_{m}$ and $b_{m}$ are, respectively,
the coefficients of the regular and irregular solutions.

In this point, we apply the boundary condition in Eq. \eqref{eq:bc}.
Doing this, one finds the following relation between the coefficients
$a_{m}$ and $b_{m}$
\begin{equation}
  \lambda_{m}\gamma^{|j|}=
  \frac{b_{m}}{a_{m}}
  \left[
    1+\frac{\lambda_{m} k^{2}}{4(1-|j|)}
    \lim_{r\rightarrow 0^{+}}r^{2-2|j|}
  \right].
  \label{eq:coef_rel_1}
\end{equation}
We note that $\lim_{r\rightarrow 0^{+}}r^{2-2|j|}$ diverges if
$|j|\geq 1$.
This condition implies that $b_m$ must be zero if $|j|\geq 1$
and only the  regular solution contributes to $f_{m}(r)$.
For $|j|<1$, when the operator $H_{0}$ is not self-adjoint, there arises
a contribution of the irregular solution to $f_{m}(r)$
\cite{AoP.2013.339.510,JPG.2013.40.075007,TMP.2013.175.637,
EPJC.2013.73.2548,AoP.2008.323.1280,TMP.2009.161.1503,
EPJC.2014.74.2708}.
In this manner, the contribution of the  irregular solution for the system
wave function stems from the fact that the operator $H_{0}$ is not
self-adjoint.

\begin{figure}
  \centering
  \includegraphics*[width=\columnwidth]{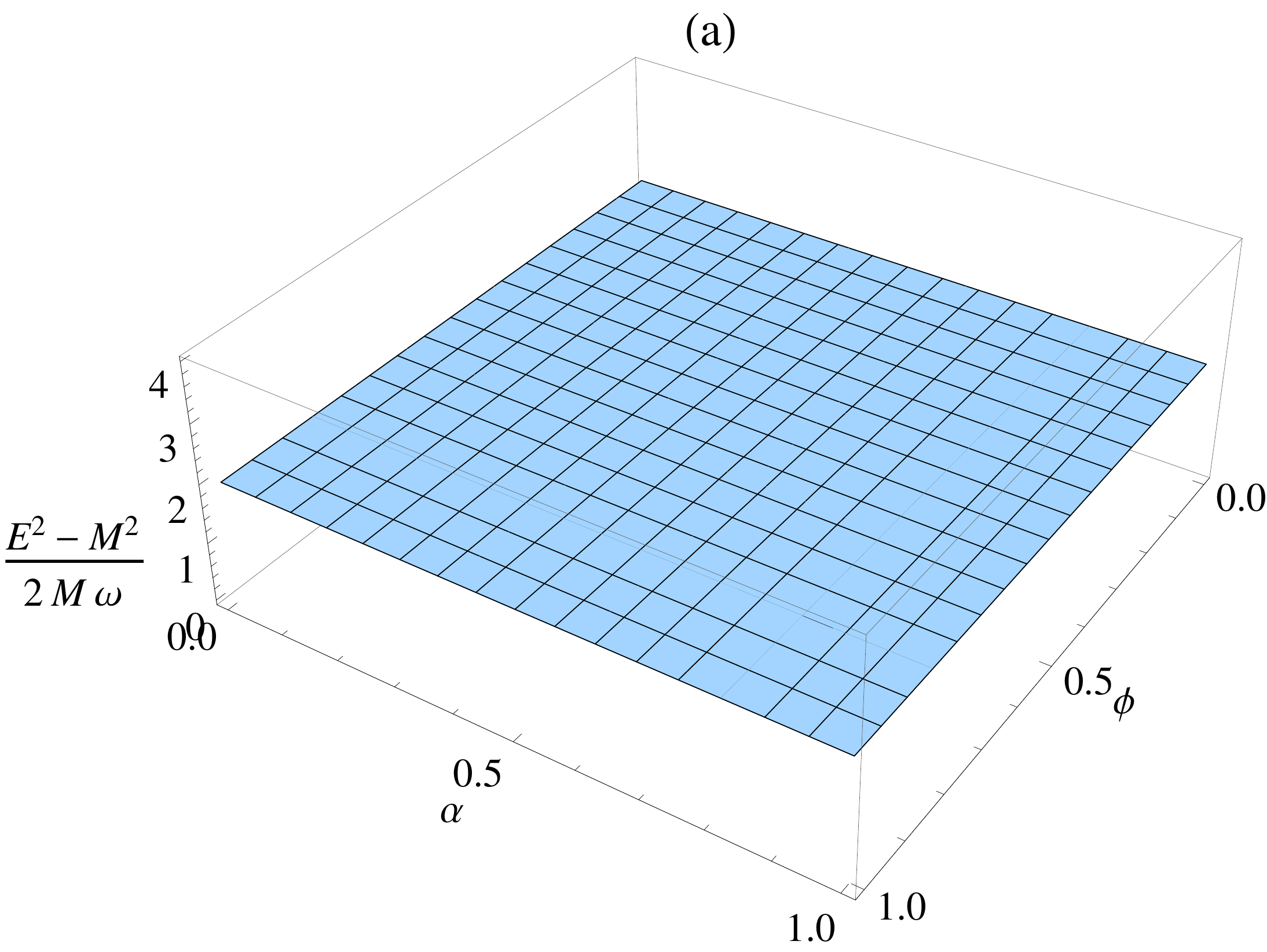}
  \includegraphics*[width=\columnwidth]{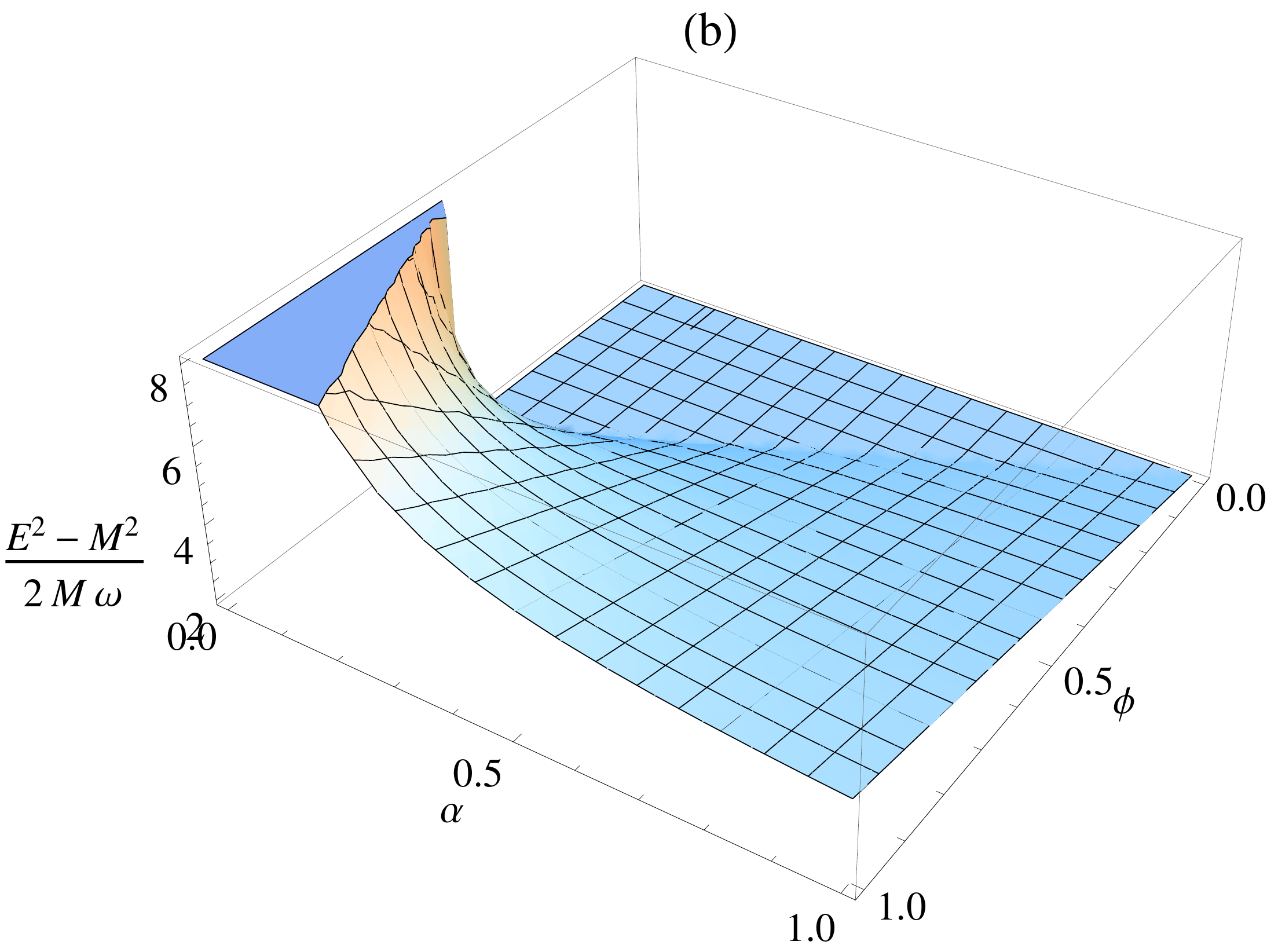}
  \caption{ \label{fig:fig1}
    (Color online) The energy eigenvalues for the regular solution
    for a 2D Dirac oscillator in the cosmic string background as a
    function of the declination $\alpha$ and the magnetic flux $\phi$
    for $n=1$, $m=1$ and for (a) $s=1$ and (b) $s=-1$.
    For convenience, we plot $(E^2-M^2)/2M\omega$ instead $E$.}
\end{figure}

For $f_{m}(r)$ be a bound state wave function, it must
vanish at large values of $r$, i.e., it must be normalizable.
So, from the asymptotic representation of the confluent hypergeometric
function, the normalizability condition is translated in
\begin{equation}
  \frac{b_{m}}{a_{m}}=
  -\frac
  {\Gamma(1+|j|)}
  {\Gamma(1-|j|)}
  \frac
  {\Gamma(d_{-})}
  {\Gamma(d_{+})}.
  \label{eq:coef_rel_2_DO}
\end{equation}
From Eq. \eqref{eq:coef_rel_1}, for $|j|<1$ we have
${b_{m}}/{a_{m}}=\lambda_{m}\gamma^{|j|}$.
Using this result in Eq. \eqref{eq:coef_rel_2_DO}, one finds
\begin{equation}
  \frac
  {\Gamma(d_{+})}
  {\Gamma(d_{-})}
  =
  -\frac{1}{\lambda_{m}\gamma^{|j|}}
  \frac
  {\Gamma(1+|j|)}
  {\Gamma(1-|j|)}.
  \label{eq:energy_BG_DO}
\end{equation}
Equation \eqref{eq:energy_BG_DO} implicitly determines the bound state
energy for the Dirac oscillator in the cosmic string background for
different values of the self-adjoint extension parameter.
Two limiting values for the self-adjoint extension parameter deserve
some attention.
For $\lambda_{m}=0$, when the $\delta$ interaction is absent, only the
regular solution contributes for the bound state wave function.
On the other side, for $\lambda_{m}=\infty$ only the irregular solution
contribute for the bound state wave function.
For all other values of the self-adjoint extension parameter, both
regular and irregular solutions contributes for the bound state wave
function.
The energies for the limiting values are obtained from the poles of the
gamma function, namely,
\begin{equation}\label{eq:gammapoles}
  \left\{
  \begin{array}{lll}
    d_{+}= -n& \mbox{for } \lambda_{m}=0 & \mbox{(regular solution)},\\
    d_{-}= -n& \mbox{for } \lambda_{m}=\infty & \mbox{(irregular solution)},
  \end{array}
  \right.
\end{equation}
with $n$ a nonnegative integer, $n=0,1,2,\ldots$.
By manipulation of Eq. \eqref{eq:gammapoles}, we obtain
\begin{align}\label{eq:energy_ABDO}
  E= {} & \pm\left\{M^{2} + 2M\omega
  \left[
    2n\pm\frac{1}{\alpha}
    \Big|m+\phi +\frac{1-\alpha}{2}s\Big|
  \right.\right.
  \nonumber \\
  & \left.
  \left.
    -\frac{s}{\alpha}
    \left(m+\phi+\frac{1-\alpha}{2}s\right)
  \right]\right\}^{1/2}.
\end{align}
In particular, it should be noted that for the case when
$|j|\geq 1$ or when the $\delta$ interaction is absent, only the
regular solution contributes for the bound state wave function
($b_m=0$), and the energy is given by Eq. \eqref{eq:energy_ABDO} using
the plus sign.
Note that, for $\alpha=1$ (flat space) and $\phi=0$ (no magnetic flux),
Eq. \eqref{eq:energy_ABDO} coincides with the energy found for the usual
2D Dirac oscillator (cf. Eq. \eqref{eq:energy2ddof} in
\ref{sec:appendixC}).
Without loss of generality, let us suppose $0<\phi<1$
\cite{AoP.1983.146.1,LMP.1998.43.43,JPA.2010.43.354011}.
In this interval, and recalling that we are interested in the case where
$0<\alpha\leq 1$, another interesting feature is present in the energy
eigenvalues.
For the regular solution, the eigenvalues are independent of $m$, $\phi$
and $\alpha$ for $s=1$.
This situation is shown in Fig. \ref{fig:fig1}(a) for $n=1$ and $m=1$.
However, this independence is absent for $s=-1$, as shown in
Fig. \ref{fig:fig1}(b) for $n=1$ and $m=1$.
In the other hand, for the irregular solution, the eigenvalues are
independent of $m$, $\phi$ and
$\alpha$ for $s=-1$ and dependent for $s=1$.
Also, for $s=-1$, decreasing the value of $\alpha$, the energy increase
as an effect of the quantum localization.

The unnormalized bound state wave functions for our problem are
\begin{align}\label{eq:wave_func}
  f_{m}(r)
  = {} &
  \rho^{\pm|j|/2}e^{-\rho/2}\;
  F\left(-n,1\pm|j|,\rho\right).
\end{align}

The self-adjoint extension is related with the presence of
the $\delta$ interaction.
In this manner, the self-adjoint extension parameter must be related
with the $\delta$ interaction coupling constant $\phi s/\alpha$.
In fact, as shown in Refs. \cite{PRD.2012.85.041701,AoP.2013.339.510}
(see also Refs.  \cite{PLB.2013.719.467,JPG.2013.40.075007}),
from the regularization of the $\delta$ interaction, it is possible to
find such a relationship.
Using the regularization method, one obtains the following equation for
the bound state energy:
\begin{equation}
  \label{eq:energy_KS_DO}
  \frac
  {\Gamma(d_{+})}
  {\Gamma(d_{-})}=
  -\frac{1}{r_{0}^{2|j|}}
  \left(
    \frac
    {\phi s+\alpha |j|}
    {\phi s-\alpha |j|}
  \right)
  \frac{1}{\gamma^{|j|}}
  \frac
  {\Gamma\left(1+|j|\right)}
  {\Gamma\left(1-|j|\right)}.
\end{equation}
By comparing Eqs. \eqref{eq:energy_BG_DO} and
\eqref{eq:energy_KS_DO}, this relation is found to be
\begin{equation}
  \label{eq:saeparam}
  \frac{1}{\lambda_{m}}
  =
  \frac{1}{r_{0}^{2|j|}}
  \left(
    \frac
    {\phi s+\alpha |j|}
    {\phi s-\alpha |j|}
  \right)
\end{equation}
where $r_0$ is a very small radius which comes from the
$\delta$ regularization \cite{PRD.2012.85.041701,AoP.2013.339.510}.

\section{Nonrelativisitic limit}
We shall now take the nonrelativistic limit of Eq. \eqref{eq:edod}.
Using $E=M+\mathcal{E}$ with $M \gg \mathcal{E}$, we obtain

\begin{equation}\label{eq:pauliosnr}
   2 M \mathcal{E}\psi = H \psi.
\end{equation}
Performing the same steps as for the relativistic case, one obtains the
shifted energy levels (cf. \ref{sec:appendixC})
\begin{align}\label{eq:energy_ABDO_nr}
   \mathcal{E}+\omega = {} &
  \left[
    1+2n\pm\frac{1}{\alpha}
    \Big|m+\phi +\frac{1-\alpha}{2}s\Big|
  \right.
  \nonumber \\
  & \left.
    -\frac{s}{\alpha}
    \left(m+\phi+\frac{1-\alpha}{2}s\right)
  \right]\omega.
\end{align}
In this equation, the $+$ ($-$) sign is for $\lambda_{m}=0$
($\lambda_{m}=\infty$) when one has regular (irregular) solution.
We note that the energy in Eq. \eqref{eq:energy_ABDO_nr} corresponds to
equation (54) of Ref. \cite{JMP.1995.36.5453} (cf. also
Eq. \eqref{eq:nrenergy2ddof} in  \ref{sec:appendixC}) with two
additional contributions, the spin-orbit coupling and the spin
connection.

\section{Conclusions}
\label{sec:conclusion}

In this contribution, we have addressed the Dirac oscillator interacting
with a topological defect and in the presence of the Aharonov-Bohm
potential.
This system has been studied in Ref.  \cite{PRA.2011.84.32109}.
However, the authors do not take into account the effects of spin.
In other words, the term proportional to the $\delta$ interaction was
discarded, by considering only the regular solution of the problem.
The presence of this term has direct implications in the energy spectrum
and wave functions of the oscillator.
The correct approach to this problem must include spin effects, which
are explicitly manifested by the spin-orbit coupling term, and, so we
have a complete description for the dynamics of the 2D Dirac oscillator.
We consider the self-adjoint extension method and show that the
spin-orbit coupling term, which results in a $\delta$ interaction,
cannot be dropped from the Hamiltonian. 
Although being singular at the origin, this term reveals that both
regular and irregular solutions contribute for the bound state wave
function and, consequently, for the energy spectrum.
Expressions for the bound states energy for different values of the
self-adjoint extension parameter were obtained.
For two specific values for the self-adjoint extension parameter, i.e.,
$\lambda_{m}=0$ and $\lambda_{m}=\infty$, the bound state energies are
given explicitly in Eq. \eqref{eq:energy_ABDO}.
We also verified that, for the flat space ($\alpha=1$) and no magnetic
flux ($\phi=0$), the results of the usual 2D Dirac oscillator are
recovered.

\appendix

\section{Covariancy of $\gamma$ matrices in the curved
spacetime}
\label{sec:appendixA}

In this appendix, we give the details of the calculation of the
covariant derivative of $\gamma^{\mu}$ in Eq. \eqref{eq:diracmathe}.
As a consequence of the tetrad postulate
\begin{equation}\label{eq:cvd}
  \nabla_{\mu}e_{\nu}^{\bar{a}}\equiv
  \partial_{\mu}e_{\nu}^{\bar{a}}
  -e_{\sigma}^{\bar{a}}\Gamma_{\mu \nu}^{\sigma}
  + \omega_{\mu \bar{b}}^{\bar{a}}e_{\nu}^{\bar{b}}=0,
\end{equation}
together with the condition
\begin{equation}
  \left[ \gamma^{\bar{a}},\Gamma_{\mu}\right] =
  \omega_{\mu \bar{b}}^{\bar{a}}\gamma^{\bar{b}},  \label{eq:concv}
\end{equation}
we have
\begin{equation}
  \nabla_{\mu}\gamma^{\lambda}=\partial_{\mu}\gamma^{\lambda}+
  \Gamma_{\mu \nu}^{\lambda}\gamma^{\nu}+
  \left[ \gamma^{\lambda},\Gamma_{\mu}\right] =0.  \label{eq:cvtly}
\end{equation}
To check this, first we need to find the relevant Christoffel symbols.
They are found to be
\begin{align}
  \Gamma_{\varphi \varphi}^{r} = {} &-\alpha^{2}r,\label{eq:syba} \\
  \Gamma_{r\varphi}^{\varphi}  = {} &\Gamma_{\varphi r}^{\varphi}=\frac{1}{r}.
\label{eq:sybb}
\end{align}
Moreover, we also make use of Eqs. \eqref{eq:ga0}-\eqref{eq:ga2} and
Eq. \eqref{eq:cvtly} and then calculate for each
$\left(\mu,\lambda=t,r,\varphi \right)$.
Then, for $\mu =t$, we write7
\begin{equation}
  \nabla_{t}\gamma^{\lambda}=\partial_{t}\gamma^{\lambda}
  +\Gamma_{t\nu}^{\lambda}\gamma^{\nu}
  +\left[\gamma^{\lambda},\Gamma_{t}\right]
.\qquad (\mu=t)
\end{equation}
Since
$\Gamma_{t\nu}^{t}=\Gamma_{t}=\partial_{t}\gamma^{t}=0$,
$\Gamma_{t\nu}^{r}=\partial_{t}\gamma^{r}=0$ and
$\Gamma_{t\nu}^{\varphi}=\partial_{t}\gamma^{\varphi}=0$,
it follows that
\begin{align}
  \nabla_{t}\gamma^{t} = {}
  & \partial_{t}\gamma^{t}+\Gamma_{t\nu}^{t}\gamma^{\nu}
    +\left[ \gamma^{t},\Gamma_{t}\right] =0,\quad (\lambda=t)\\
  \nabla_{t}\gamma^{r}  = {}
  & \partial_{t}\gamma^{r}+\Gamma_{t\nu}^{r}\gamma^{\nu}
    +\left[ \gamma^{r},\Gamma_{t}\right] =0,\quad (\lambda=r)  \\
  \nabla_{t}\gamma^{\varphi} = {}
  &\partial_{t}\gamma^{\varphi}
    +\Gamma_{t\nu}^{\varphi}\gamma^{\nu}+
    \left[ \gamma^{\varphi},\Gamma_{t}\right] =0,\quad(\lambda =\varphi)
\end{align}
so that
\begin{equation}
  \nabla_{t}\gamma^{\lambda}=\nabla_{t}\gamma^{t}+\nabla_{t}\gamma^{r}
  +\nabla_{t}\gamma^{\varphi}=0.\quad(\lambda=t,r,\varphi)
  \label{eq:Gt}
\end{equation}
Next, we have
\begin{equation}
  \nabla_{r}\gamma^{\lambda}=\partial_{r}\gamma^{\lambda}
  +\Gamma_{r\nu}^{\lambda}\gamma^{\nu}
  +\left[ \gamma^{\lambda},\Gamma_{r}\right],\quad(\mu=r)
\end{equation}
and by using
$\Gamma_{r\varphi}^{\varphi}=1/r$ (with $\nu =\varphi$),
$\Gamma_{r\nu}^{t}=\Gamma_{r\nu}^{r}=0$,
$\partial_{t}\gamma^{t}=\partial_{r}\gamma^{r}=0$, we get
\begin{align}
  \nabla_{r}\gamma^{t}  = {}
  &\partial_{r}\gamma^{t}+\Gamma_{r\nu}^{t}\gamma^{\nu}
    +\left[ \gamma^{t},\Gamma_{r}\right] =0,\quad\left( \lambda =t\right)\\
  \nabla_{r}\gamma^{r} = {}
  &\partial_{r}\gamma^{r}+\Gamma_{r\nu}^{r}\gamma^{\nu}
    +\left[ \gamma^{r},\Gamma_{r}\right] =0,\quad\left( \lambda =r\right)\\
  \nabla_{r}\gamma^{\varphi}  = {}
  & \partial_{r}\gamma^{\varphi}+\Gamma_{r\nu}^{\varphi}\gamma^{\nu}
    +\left[ \gamma^{\varphi},\Gamma_{r}\right] ,  \nonumber \\
  = {} &\frac{1}{\alpha r^{2}}\left( \gamma^{\bar{1}}\sin \varphi
         -\gamma^{\bar{2}}\cos \varphi \right)   \nonumber \\
  & +\frac{1}{\alpha r^{2}}\left( -\gamma^{\bar{1}}\sin \varphi
    +\gamma^{\bar{2}}\cos \varphi \right) ,  \nonumber \\
  = {} & 0,\quad\left( \lambda =\varphi \right)
\end{align}
and consequently
\begin{equation}
  \nabla_{r}\gamma^{\lambda}=\nabla_{r}\gamma^{t}
  +\nabla_{r}\gamma^{r}+\nabla_{r}\gamma^{\varphi}=0.
  \quad\left( \lambda =t,r,\varphi \right)
  \label{eq:Gr}
\end{equation}
Now, for $\mu =\varphi $, we write
\begin{equation}
  \nabla_{\varphi}\gamma^{\lambda}=\partial_{\varphi}\gamma^{\lambda}
  +\Gamma_{\varphi \nu}^{\lambda}\gamma^{\nu}
  +\left[ \gamma^{\lambda},\Gamma_{\varphi}\right] ,
  \quad\left( \mu =\varphi \right)   \label{eq:lmr}
\end{equation}
and again, since
$\partial_{\varphi}\gamma^{t}=\Gamma_{\varphi\nu}^{t}=0$,
and using Eqs. \eqref{eq:syba} and \eqref{eq:sybb}, we
have
\begin{align}
  \nabla_{\varphi}\gamma^{t}
  = {} & \partial_{\varphi}\gamma^{t}+\Gamma_{\varphi \nu}^{t}\gamma^{\nu}
         -\left[\Gamma_{\varphi},\gamma^{t}\right] ,  \nonumber \\
  = {} & \frac{1}{2}(1-\alpha)\gamma_{\bar{1}}\gamma_{\bar{2}}\gamma^{t}
         -\frac{1}{2}(1-\alpha)\gamma^{t}\gamma_{\bar{1}}\gamma_{\bar{2}},
         \nonumber \\
  = {} & 0, \quad (\lambda=t).
\end{align}
\begin{align}
  \nabla_{\varphi}\gamma^{r}
  = {} & \partial_{\varphi}\gamma^{r}+\Gamma_{\varphi\nu}^{r}\gamma^{\nu}
         -\left[\Gamma_{\varphi},\gamma^{r}\right],  \nonumber \\
  = {} &-(1-\alpha) \gamma^{\bar{2}}\cos \varphi
         -(1-\alpha) \gamma^{\bar{1}}\sin \varphi   \nonumber \\
       & +\left( 1-\alpha \right) \gamma^{\bar{2}}\cos \varphi -\left( 1-\alpha
\right) \gamma^{\bar{1}}\sin \varphi ,  \nonumber \\
= {} & 0,\quad (\lambda=r).
\end{align}
\begin{align}
  \nabla_{\varphi}\gamma^{\varphi}
  = {} &\partial_{\varphi}\gamma^{\varphi}
         +\Gamma_{\varphi \nu}^{\varphi}\gamma^{\nu}
         -\left[ \Gamma_{\varphi},\gamma^{\varphi}\right] ,  \nonumber \\
  = {} &-\frac{1}{\alpha r}(1-\alpha)
         \left(
         \gamma^{\bar{1}}\cos\varphi +\gamma^{\bar{2}}\sin \varphi
         \right)   \nonumber \\
       &+\frac{1}{\alpha r}(1-\alpha)
         \left(
         \gamma^{\bar{1}}\cos\varphi +\gamma^{\bar{2}}\sin \varphi
         \right) ,  \nonumber \\
  = {} &0, \quad (\lambda=\varphi),
\end{align}
so that
\begin{equation}
  \nabla_{\varphi}\gamma^{\lambda}=
  \nabla_{\varphi}\gamma^{t}+\nabla_{\varphi}\gamma^{r}
  +\nabla_{\varphi}\gamma^{\varphi}=0.\quad (\lambda=t,r,\varphi).
  \label{eq:Gf}
\end{equation}
Therefore, Eqs. \eqref{eq:Gt}, \eqref{eq:Gr} and \eqref{eq:Gf}, imply
that
\begin{equation}
\nabla_{\mu}\gamma^{\lambda}=0.\quad (\mu,\lambda=t,r,\varphi)
\end{equation}
Thus, we have verified that the matrices $\gamma^{\mu}$, in the basis
tetrad given in Eq. \eqref{eq:diracmathe}, are covariantly constant.

\section{Derivation of the spin connection}
\label{sec:appendixB}

The spinor affine connection in Eq. \eqref{eq:conn} can be written more
explicitly as
\begin{align}
  \Gamma_{t} = {} &
  \frac{1}{4}\omega_{t}^{\bar{0}\bar{1}}
  \left[\gamma_{\bar{0}},\gamma_{\bar{1}}\right]
  +\frac{1}{4}\omega_{t}^{\bar{0}\bar{2}}
  \left[\gamma_{\bar{0}},\gamma_{\bar{2}}\right]
  +\frac{1}{4}\omega_{t}^{\bar{1}\bar{2}}
  \left[\gamma_{\bar{1}},\gamma_{\bar{2}}\right], \\
  \Gamma_{r} = {} &
  \frac{1}{4}\omega_{r}^{\bar{0}\bar{1}}
  \left[\gamma_{\bar{0}},\gamma_{\bar{1}}\right]
  +\frac{1}{4}\omega_{r}^{\bar{0}\bar{2}}
  \left[\gamma_{\bar{0}},\gamma_{\bar{2}}\right]
  +\frac{1}{4}\omega_{r}^{\bar{1}\bar{2}}
  \left[\gamma_{\bar{1}},\gamma_{\bar{2}}\right], \\
  \Gamma_{\varphi} = {} &
  \frac{1}{4}\omega_{\varphi}^{\bar{0}\bar{1}}
  \left[ \gamma_{\bar{0}},\gamma_{\bar{1}}\right]
  +\frac{1}{4}\omega_{\varphi}^{\bar{0}\bar{2}}
  \left[ \gamma_{\bar{0}},\gamma_{\bar{2}}\right]
  +\frac{1}{4}\omega_{\varphi}^{\bar{1}\bar{2}}
  \left[ \gamma_{\bar{1}},\gamma_{\bar{2}}\right] , \\
  \Gamma_{z} = {} &
  \frac{1}{4}\omega_{z}^{\bar{0}\bar{1}}
  \left[ \gamma_{\bar{0}},\gamma_{\bar{1}}\right]
  +\frac{1}{4}\omega_{z}^{\bar{0}\bar{2}}
  \left[\gamma_{\bar{0}},\gamma_{\bar{2}}\right]
  +\frac{1}{4}\omega_{z}^{\bar{1}  \bar{2}}
  \left[ \gamma_{\bar{1}},\gamma_{\bar{2}}\right].
\end{align}
In order to calculate the spin connection $\omega_{\mu \bar{a}\bar{b}}$
in Eq. \eqref{spincn}, we use the Christoffel symbols given in Eqs.
\eqref{eq:syba} and \eqref{eq:sybb}.
Since
$\omega_{t}^{\bar{0}\bar{1}}=\omega_{t}^{\bar{0}\bar{2}}=
\omega_{t}^{\bar{1}\bar{2}}=0$,
$\omega_{r}^{\bar{0}\bar{1}}=\omega_{r}^{\bar{0}\bar{2}}=
\omega_{r}^{\bar{1}\bar{2}}=0$ and
$\omega_{z}^{\bar{0}\bar{1}}=\omega_{z}^{\bar{0}\bar{2}}=
\omega_{z}^{\bar{1}\bar{2}}=0$, we find
$\Gamma_{t}=\Gamma_{r}=\Gamma_{z}=0$.
Furthermore, we also can verify that
$\omega_{\varphi}^{\bar{0}\bar{1}}=\omega_{\varphi}^{\bar{0}\bar{2}}=0$.
As a result, we have
\begin{equation}
  \Gamma_{\varphi \nu}^{\mu}\neq 0,
\end{equation}
for $\mu=\varphi$, $\nu=r$ and $\mu=r$, $\nu=\varphi$ .
Thus, the only contribution for the spin connection is obtained from
\begin{equation}
  \omega_{\varphi}^{\bar{1}\bar{2}}=
  e_{\mu}^{\bar{1}}e^{\nu \bar{2}}\Gamma_{\varphi \nu}^{\mu}
  -e^{\nu \bar{2}}\partial_{\varphi}e_{\nu}^{\bar{1}},
\end{equation}
which gives
\begin{align}
  \omega_{\varphi}^{\bar{1}\bar{2}} = {} &
  -\alpha r\sin \varphi
  \sin \varphi \frac{1}{r}+
  \cos \varphi
  \frac{1}{\alpha r}\cos \varphi
  \left(-\alpha^{2}r\right)  \nonumber \\
  &
  -\sin \varphi \partial_{\varphi}\cos \varphi
  -\frac{1}{\alpha r}\cos \varphi
  \partial_{\varphi}\left(-\alpha r\sin \varphi \right), \nonumber \\
  = {} & 1-\alpha .
\end{align}
Thus, for the specific basis tetrad \eqref{eq:tetrad}, the connection is
found to be
\begin{equation}
  \boldsymbol{\Gamma}=\left( 0,0,\Gamma_{\varphi},0\right) ,
\end{equation}
with the non-vanishing element given as
\begin{equation}
  \Gamma_{\varphi}=-\frac{1}{4}\left( 1-\alpha \right)
  \left[ \gamma_{\bar{1}},\gamma_{\bar{2}}\right] =
  -\frac{1}{2}\left( 1-\alpha \right) \gamma_{\bar{1}}\gamma_{\bar{2}}.
\end{equation}

\section{2D Dirac oscillator}
\label{sec:appendixC}

In this appendix, we briefly discuss the usual 2D Dirac oscillator.
We mention that although fully equivalent, the present construction is
slightly different from the previous construction one in the literature
\cite{EPL.2014.108.30003}.
Let us consider Eq. \eqref{eq:dirac2} with
$\boldsymbol{\pi}=\mathbf{p}-iM\omega\tilde{\beta}\mathbf{r}$.
By using the representation for the $\tilde{\gamma}$ matrices in
Eq. \eqref{eq:newmatrices}, we are left with
\begin{equation}\label{eq:diracsigmado}
[\sigma^{1}\pi_{1}+s\sigma^{2}\pi_{2}+\sigma^{3}M-E]\psi=0,
\end{equation}
with $\pi_{i}=p_{i}-iM\omega \sigma^{3}r_{i}$.
By squaring Eq. \eqref{eq:diracsigmado}, one obtains
\begin{equation}\label{eq:2ddiracoscillator}
  \left[
    p^2+M^2\omega^2r^2-2M\omega(\sigma^{3}+s L_{3})
  \right]\psi=
  (E^2-M^{2})\psi.
\end{equation}
Equation \eqref{eq:2ddiracoscillator}, restoring the factors $\hbar$ and
$c$, in terms of components, provides
\begin{subequations}\label{eq:2ddiracoscillatorcomp}
\begin{equation}
  2Mc^{2}\left[
    H_{\rm{ho}}^{\rm 2D}-\hbar\omega-s\omega L_{3}
  \right]\psi_1=
  (E^2-M^{2}c^{4})\psi_1,
\end{equation}
\begin{equation}
  2Mc^{2}\left[
    H_{\rm ho}^{\rm 2D}+\hbar\omega-s\omega L_{3}
  \right]\psi_2=\\
  (E^2-M^{2}c^{4})\psi_2,
\end{equation}
\end{subequations}
where
\begin{equation}
H_{\rm ho}^{\rm 2D}=\frac{p^2}{2M}+\frac{1}{2}M\omega^2r^2.
\end{equation}
Equation \eqref{eq:2ddiracoscillatorcomp} for $s=1$
agreed with the expressions found in Eq. (A2) of
Ref. \cite{PRA.2008.77.033832} and Eqs. (9) and (22) of
Ref. \cite{MPLA.2004.19.2147}.
Using the ansatz in Eq. \eqref{eq:soloc} the energy eigenvalues are
determined:
\begin{equation}\label{eq:energy2ddof}
E=\pm\sqrt{M^{2}+2M\omega\left(2n+|m|-sm\right)},
\end{equation}
showing that the energy eigenvalues are spin dependent.
It should be noted that for $s=1$ ($s=-1$) and $m>0$ ($m<0$) the energy
eigenvalues are independent of the quantum number $m$.

From Eq. \eqref{eq:2ddiracoscillator}, in the nonrelativistic limit
$E=M+\mathcal{E}$ with $M \gg \mathcal{E}$, we have
\begin{equation}\label{eq:diraconr}
  \left[
    H_{\rm{ho}}^{\rm 2D}
   -\omega(\sigma^{3}+s L_{3})
  \right]\psi=
  \mathcal{E}\psi.
\end{equation}
The first term on the left side of Eq. \eqref{eq:diraconr} is the
Hamiltonian of the nonrelativistic circular harmonic oscillator
\cite{Book.1999.Flugge}, explaining why this system is called the Dirac 
oscillator.
The second term is a constant which shifts all energy levels.
The last term is the spin-orbit coupling, which (restoring the factor
$\hbar$) is of strength $\omega/\hbar$.
Summarizing, the nonrelativistic limit of the 2D Dirac oscillator is the
circular harmonic oscillator with a strong spin-orbit coupling term with
all levels shifted by the factor $\omega$.
Indeed, the shifted energy levels are
\begin{equation}\label{eq:nrenergy2ddof}
\mathcal{E}+\omega=(1+2n+|m|-sm)\omega.
\end{equation}
As for the relativistic case, for $s=1$ ($s=-1$) and $m>0$ ($m<0$) the
energy eigenvalues are independent of the quantum number $m$.

\section*{Acknowledgments}
We would like to thank R. Casana and L. R. B. Castro for fruitful
discussions.
This work was supported by the
Funda\c{c}\~{a}o Arauc\'{a}ria (Grant No. 205/2013 (PPP) and
No. 484/2014 (PQ)),
and the Conselho Nacional de Desenvolvimento
Cient\'{i}fico e Tecnol\'{o}gico (Grants No. 482015/2013-6 (Universal),
No. 306068/2013-3 (PQ)) and FAPEMA (Grant No. 00845/13).
Finally, we acknowledge some suggestions made by the anonymous referees
in order to improve the present work.

\bibliographystyle{spphys}

\end{document}